\documentclass[letterpaper,twocolumn]{jpsj3}
\usepackage{graphicx}

\title{Inverse Mellin Transformation of Continuous Singular Value Decomposition: A Route to Holographic Renormalization}

\author{Hiroaki Matsueda\thanks{matsueda@sendai-nct.ac.jp}}
\inst{Sendai National College of Technology, Sendai 989-3128, Japan} 

\recdate{\today}

\abst{
We examine holographic renormalization by the singular value decomposition (SVD) of matrix data generated by the Monte Carlo snapshot of the 2D classical Ising model at criticality. To take the continuous limit of the SVD enables us to find the mathematical form of each SVD component by the inverse Mellin transformation as well as the power-law behavior of the SVD spectrum. We find that each SVD component is characterized by the two-point spin correlator with a finite correlation length. Then, the continuous limit of the decomposition index in the SVD corresponds to the inverse of the correlation length. These features strongly suggest that the SVD contains mathematical structure the same as the holographic renormalization.
}

\begin{document}

\maketitle

\section{Introduction}

Quantum entanglement and holography are known to be complementary concepts in recent development of renormalization group (RG) theory and information physics. The former can be characterized by the singular value decomposition (SVD), and the SVD offers indispensable techniques for the density matrix renormalization group (DMRG) and the tensor-network variational methods in the condensed matter physics~\cite{White,Verstraete}. On the other hand, the relation between the SVD and the latter concept remains to be an open question. Although some relationship has been pointed out by recent works for the snapshot entropy in the 2D classical spin systems~\cite{Matsueda1,Imura,Matsueda2,Matsueda3,Matsueda4}, its description is in just an elementary level. Thus, its mathematically precise formulation is required. Furthermore, this type of study has not been done yet in quantum systems. If the study has been done (that would is possible by using the DMRG in the 1D case), we may think of better understanding of recent important topics such as the anti-de Sitter space/conformal field theory (AdS/CFT) correspondence in the string theory~\cite{Maldacena} and the multiscale entanglement renormalization ansatz (MERA) in the condensed matter physics~\cite{Vidal}.

One technical difficulty is that the SVD itself is a discrete decomposition. Thus it is hard to perform its inverse transformation unlike several integral transformations. As I have already proposed in the previous papers~\cite{Matsueda1,Matsueda2,Matsueda3,Matsueda4}, we can simply consider that the data decomposed by the SVD are stored into a space with one-higher dimension, and we can notice that the length scale in the data space changes along the decomposition index. However, it is not obvious whether the continuous limit of the decomposition index really corresponds to the radial axis of the AdS or a flow parameter of the holographic RG. There is no well-grounded formulation in the present stage. Here, we examine some mathematical properties of the continuous limit of the SVD, and we call this as the continuous SVD (CSVD) in short. By using such mathematical formulation, we can examine their relationship to the holography.

The point to resolve the abovementioned difficulty is to notice the presence of the universal scaling formula for the singular value spectrum. For instance, let us consider the SVD of the spin configuration data in the 2D classical Ising model at criticality. In the author's previous works~\cite{Matsueda3,Matsueda4}, it has been found that the scaling is associated with the power law behavior of the two-point correlator. In this case, we can introduce the inverse transformation of the CSVD. This is because if the power-law scaling exists the transformation has clear mathematical meaning. This is called as the inverse Mellin transformation that plays a crucial role on the data decomposition of the scale-invariant system. In the string theory side also, it has recently been discussed by using the Mellin representation that the CFT correlators are rewritten as AdS scattering amplitudes~\cite{Mack,Penedones,Paulos,Fitspatrick}. Therefore, I believe that the present work is also an important complementary work to these recent works. When the inverse Mellin transformation of the CSVD exists, we can examine how the spin correlation length changes as we change the continuous parameter. The examination of such parameter flow is the purpose of this paper. We will conclude that the continuous limit of the decomposition index corresponds to the inverse correlation length, and thus the SVD generates the series of the holographic RG data.

\section{Formulation}

\subsection{Continuous Singular Value Decomposition at the Critical Point}

Let us start with the Ising model on the 2D square lattice
\begin{eqnarray}
H=-J\sum_{<ij>}\sigma_{i}\sigma_{j},
\end{eqnarray}
where we consider the ferromagnetic coupling $J>0$ and the Ising spin at site $i$ takes $\sigma_{i}=\pm 1$. Here we take a snapshot of the spin configuration at $T_{c}$ by the cluster Monte Carlo simulations, which is represented as a $L_{x}\times L_{y}$ matrix $M(x,y)$. We apply SVD to decompose $M$ into the following form
\begin{eqnarray}
M(x,y)=\sum_{n=1}^{L}U_{n}(x)\sqrt{\lambda_{n}}V_{n}(y),
\end{eqnarray}
where we denote $L={\rm min}\left(L_{x},L_{y}\right)$, $U_{n}(x)$ and $V_{n}(y)$ are column unitary matrices
\begin{eqnarray}
\sum_{x}U_{n}(x)U_{n^{\prime}}(x)=\sum_{y}V_{n}(y)V_{n^{\prime}}(y)=\delta_{nn^{\prime}},
\end{eqnarray}
and this relation leads to
\begin{eqnarray}
\sum_{x,y}M^{2}(x,y)=\sum_{n}\lambda_{n}.
\end{eqnarray}
We call $U_{n}(x)V_{n}(y)$ as the $n$-th layer data, and find that $U_{n}(x)V_{n}(y)$ represents the $n$-th largest cluster data in the snapshot when we label the data as $\lambda_{1}\ge\lambda_{2}\ge\cdots\lambda_{L}$. We also define the partial density matrix as
\begin{eqnarray}
\rho(x,x^{\prime}) = \sum_{y}M(x,y)M(x^{\prime},y) = \sum_{n=1}^{L}U_{n}(x)\lambda_{n}U_{n}(x^{\prime}).
\end{eqnarray}
In this work, we focus on this quantity. There is a symmetry on $\rho(x,x^{\prime})$ associated with the exchange between $x$ and $x^{\prime}$, $\rho(x,x^{\prime})=\rho(x^{\prime},x)$. As an important argument, it has been shown that this is essentially equal to the two-point correlator $C(l)$ with $l=|x-x^{\prime}|$ due to self-averaging~\cite{Imura}. More precisely, a single snapshot does not have the translational invariance, and we need to take the sample average by a set of various snapshots to reduce the sample dependence on $\rho(x,x^{\prime})$. Fortunately, it has been found numerically that the asymptotic behavior of the SVD spectrum is independent of the sample difference in the large-$L$ cases~\cite{Matsueda3,Matsueda4}, and we expect that the average has a simple form. In this case, we can take
\begin{eqnarray}
C(|x-x^{\prime}|) = \left<\rho(x,x^{\prime})\right> = \sum_{n=1}^{L}\left<U_{n}(x)U_{n}(x^{\prime})\right>\lambda_{n},
\end{eqnarray}
with the angle bracket representing the sample avarage. Thus the basic structure of the SVD spectrum still remains. Hereafter we denote $\left<U_{n}(x)U_{n}(x^{\prime})\right>$ as $R_{n}(|x-x^{\prime}|)$.

Now, we introduce the continuous limit of the abovementioned quantities, called CSVD in this paper. The most straightforward way is to assume the existence of the following decomposition for a two-parameter function $M(x,y)$ ($0\le x\le\infty,0\le y\le\infty$)
\begin{eqnarray}
M(x,y)=\int_{0}^{\infty}dz U(z,x)\sqrt{\lambda(z)}V(z,y),
\end{eqnarray}
with the unitary conditions
\begin{eqnarray}
\int dx U(z,x)U(z^{\prime},x)=\int dy V(z,y)V(z^{\prime},y) \nonumber \\
=\delta(z-z^{\prime}),
\end{eqnarray}
and
\begin{eqnarray}
\iint dxdy M^{2}(x,y)=\int dz\lambda(z).
\end{eqnarray}
Here, $\lambda(z)$ is a monotone decreasing function, and is none-negative. The continuous limit of the SVD index $n$ corresponds to the parameter $z$. If this assumption is correct, the density matrix and the correlation function are given by
\begin{eqnarray}
\rho(x,x^{\prime})=\int_{0}^{\infty}dz U(z,x)\lambda(z)U(z,x^{\prime}),
\end{eqnarray}
and
\begin{eqnarray}
C(l) = \rho(l) = \int_{0}^{\infty}dz R(l,z)\lambda(z),
\end{eqnarray}
\begin{eqnarray}
R(l,z) = \iint dxdx^{\prime}U(z,x)U(z,x^{\prime})\delta(l-|x-x^{\prime}|), \label{eq12}
\end{eqnarray}
where it is not necessary to take the sample average due to the perfect self-averaging in the thermodynamic limit. Hereafter, we need to examine the uniqueness of the abovementioned continuous decomposition.

In the previous works~\cite{Matsueda3,Matsueda4}, it was found that the SVD spectrum near $T_{c}$ behaves as an algebraic function $\lambda_{n}=\lambda_{1}n^{-\Delta}$ with an exponent $\Delta$. Thus the continuous limit $\lambda(z)$ can be given by
\begin{eqnarray}
\lambda_{n}=\frac{\lambda_{1}}{n^{\Delta}} \rightarrow \lambda(z)=\frac{\lambda_{1}}{z^{\Delta}}, \label{eq4}
\end{eqnarray}
We also denote $R_{n}(l)$ as $R(l,z)$, and then obtain
\begin{eqnarray}
\rho(l)=\lambda_{1}\int_{0}^{\infty}dz R(l,z)z^{-\Delta}.
\end{eqnarray}
In particular, the result in the 2D classical Ising model at criticality is given by
\begin{eqnarray}
\Delta=1-\eta,
\end{eqnarray}
with $\eta=1/4$, and we find
\begin{eqnarray}
\rho(l)=\rho_{\eta}(l)=\lambda_{1}\int_{0}^{\infty}dz R(l,z)z^{\eta-1}.
\end{eqnarray}
Hereafter we add the suffix $\eta$ to $\rho(l)$, $\rho_{\eta}(l)$, in order to emphasize the presence of the anomalous dimension $\eta$. This is because $\eta$ is an important conjugate parameter to $z$ when we consider the inverse transformation. Only one point we should be careful for is about the normalization or the bounded condition of the SVD spectrum. If we simply consider the thermodynamic limit of Eq.~(\ref{eq4}), the sum of all spectra diverges for the condition $\Delta\le 1$. According to the author's previous works, it is better to assume
\begin{eqnarray}
\lambda(z)=\frac{f(z)}{z^{1-\eta}}
\end{eqnarray}
with a decreasing function $f(z)$. Then we have
\begin{eqnarray}
\rho_{\eta}(l)&=&\int_{0}^{\infty}dz R(l,z)f(z)z^{\eta-1} \nonumber \\
&\equiv&\int_{0}^{\infty}dz {\cal R}(l,z)z^{\eta-1}, \label{eq9}
\end{eqnarray}
and the normalization condition
\begin{eqnarray}
\int_{0}^{\infty}f(z)z^{\eta-1}dz=1.
\end{eqnarray}
Going back to the definition of the CSVD and Eq.~(\ref{eq12}), their relevance is related to the presence of the unique inverse transformation of Eq.~(\ref{eq9}). Actually, we will later find that this consideration is reasonable in the present critical case. Furthermore, if there exists the unique inverse transformation, we can obtain the explicit form of ${\cal R}(l,z)$ from the definition of $\rho_{\eta}(l)$ automatically. It is noted that the integral converges when the weight function $f(z)$ is an exponentially decreasing function. Actually, we find
\begin{eqnarray}
f(z)=\frac{\beta^{\eta}}{\Gamma(\eta)}e^{-\beta z}, \label{eq11}
\end{eqnarray}
with use of the gamma function $\Gamma(\eta)$ and a constant $\beta$ to be determined so that this function matches well with the numerical result. This form reminds us with the exponential damping factor of the two-point spin correlator with a finite correlation length $\xi$ away from the critical point, when we assume $z=\xi^{-1}$ and $\beta=l$. We cannot exclude a possibility that more precise data fitting may indicate a different form of the decreasing function $f(z)$, but now we would like to construct a conceptual or phenomenological understanding of the data structure of the each SVD component. Thus, we believe that the essential result does not change even if we take an another form of the function $f(z)$. In the end of this section, we mention some extension of that form.

The point here is to regard the integral of Eq.~(\ref{eq9}) as the sum of data associated with the RG flow. In this case, the data at a particular $z$ corresponds to the correlator with a finite correlation length $\xi$. We simply imagine $z\propto\xi^{-1}$, since the small $z$-region represents a larger cluster scale in the snapshot of the 2D Ising model. In general, the Ornstein-Zernike form of the two-point correlator $C(l)$ near $T_{c}$ is given by
\begin{eqnarray}
C(l)=\frac{A}{l^{d-2+\eta}}e^{-l/\xi}=\frac{A}{l^{\eta}}e^{-l/\xi},
\end{eqnarray}
where $d$ is the spatial dimension and $A$ is an overall constant. According to this formula, it is natural to assume
\begin{eqnarray}
{\cal R}(l,z)=\frac{Ae^{-zl}}{(zl)^{\eta^{\prime}}}, \label{eq13}
\end{eqnarray}
if the radial axis $z$ represents $\xi^{-1}$ and also represents the direction of the holographic renormalization. Then, the partial density matrix or the correlator is given by
\begin{eqnarray}
\rho_{\eta}(l) = \frac{A}{l^{\eta^{\prime}}}\int_{0}^{\infty}dz e^{-zl}z^{\eta-\eta^{\prime}-1} = \frac{A}{l^{\eta}}\Gamma\left(\eta-\eta^{\prime}\right). \label{eq14}
\end{eqnarray}
We find that this form agrees well with the correlator at the critical point, although the result contains an additional regulator $\Gamma(\eta-\eta^{\prime})$. Later, we will again mention the importance of this regulator in the inverse transformation of the CSVD. We expect that the power $\eta^{\prime}$ of the algebraic decay in the expression of ${\cal R}(l,z)$ is basically equal to the original anomalous dimension $\eta$, but mathematically we should take a value silightly smaller one to $\eta$ for keeping the convergence of the gamma function ($\Gamma(0)=\infty$).

In the present stage, $\eta^{\prime}$ is a phenomenological parameter to be determined by the fitting with numerical results. If $\eta^{\prime}$ is exactly equal to $\eta$, it is even possible to eliminate the divergence by introducing the upper incomplete gamma function with the IR cut-off $z_{0}$ as
\begin{eqnarray}
\rho_{\eta}(l,z_{0}) = \frac{A}{l^{\eta}}\int_{z_{0}}^{\infty}dz e^{-zl}z^{-1} = \frac{A}{l^{\eta}}\Gamma(0,z_{0}l). \label{eq16}
\end{eqnarray}
The incomplete gamma function $\Gamma(0,z_{0}l)$ has several expansion formulae, and for instance we have 
\begin{eqnarray}
\Gamma(0,z_{0}l) &=& e^{-z_{0}l}U(1,1,z_{0}l) \nonumber \\
&=& \frac{A}{l^{\eta}}\left(-\gamma-\ln(z_{0}l)-\sum_{k=1}^{\infty}\frac{(-z_{0}l)^{k}}{k(k!)}\right), \label{eq17}
\end{eqnarray}
where $U(1,1,z_{0}l)=\int_{0}^{\infty}du e^{-u}/(z_{0}l+u)$ is the confluent hypergeometric function and $\gamma$ is the Euler constant. This calculation also produces the $l^{-\eta}$ term, but we also find the additional $l$-dependent factor $\Gamma(0,z_{0}l)$. Particularly in the hypergeometric-function representation, we find the exponential damping factor $e^{-z_{0}l}$ with finite correlation length $z_{0}=\xi_{0}^{-1}$. That is quite natural, since we have introduced the IR cut-off.

It depends on problems to decide one of which regularization is better. The latter is very straightforward except for the presence of the damping factor away from the critical point. However, when we consider the inverse transformation, it is necessary to use the former method. 
As will be discussed, the inversion is well-defined only for the former case. Then, a $\eta$-dependent regulator with poles on the complex-$\eta$ plane is necessary, and the gamma-function regulator plays a role on the presence of these poles.

We again argue that the RG flow parameter corresponds to the inverse correlation length
\begin{eqnarray}
z=\frac{1}{\xi}.
\end{eqnarray}
Therefore, the data set of the SVD is that of different length scales. The new parameter $z$ also acts as a parameter of the scale transformation. We call the condition $z\rightarrow 0$ as the boundary of the $(l,z)$-space. The result can be briefly summarized as
\begin{eqnarray}
{\cal R}\left(\left(\frac{\epsilon}{z}\right)l,z\right)={\cal R}\left(l,\epsilon\right).
\end{eqnarray}

Before going into the next step, we consider some generalization of the damping factor $f(z)$. The most general form of $f(z)$ seems to be
\begin{eqnarray}
f(z)\propto e^{-\beta z^{\kappa}},
\end{eqnarray}
where $\kappa$ is a fitting parameter, and we have already examined the case that $\kappa=1$. For instance, an Ising-spin configuration is in some sense a random fractal. In this case, $\kappa$ deviates from unity and is rather close to $\kappa=2$~\cite{CHL}. When we assume $R(l,z)=Ae^{-\beta z^{\kappa}}/(zl)^{\eta^{\prime}}$, we find
\begin{eqnarray}
\rho_{\eta}(z) &=& \lim_{\eta^{\prime}\rightarrow\eta}\frac{A}{l^{\eta^{\prime}}}\int_{0}^{\infty}dz e^{-\beta z^{\kappa}}z^{\eta-\eta^{\prime}-1} \nonumber \\
&=& \lim_{\eta^{\prime}\rightarrow\eta}\frac{A}{\kappa\beta^{(\eta-\eta^{\prime})/\kappa}l^{\eta^{\prime}}}\Gamma\left(\frac{\eta-\eta^{\prime}}{\kappa}\right),
\end{eqnarray}
or otherwise we can introduce a cut-off and then
\begin{eqnarray}
\rho_{\eta}(l,z_{0}) = \frac{A}{l^{\eta}}\int_{z_{0}}^{\infty}dz e^{-\beta z^{\kappa}}z^{-1} = \frac{A}{\kappa l^{\eta}}\Gamma(0,\beta z_{0}^{\kappa}).
\end{eqnarray}
In both cases, we obtain the $l^{\eta}$ term for $\eta^{\prime}\rightarrow\eta$. Then, we identify
\begin{eqnarray}
z^{\kappa}=\frac{1}{\xi},
\end{eqnarray}
for $\beta=l$. In this case also, the correlation length $\xi$ increases, as the RG flow proceeds by decreasing $z$.

\subsection{Inverse Mellin Transformation}

The CSVD is a kind of the integral transformation, and the transformation in Eq.~(\ref{eq9}) is called as Mellin transformation of the function ${\cal R}(l,z)$. The peculiar feature of this transformation is that the anomalous dimension itself is the transformation parameter conjugate to the scale parameter $z$. An interesting view is that this scale/Mellin space approach was also found in the string theory side, as already mentioned in the introduction~\cite{Mack,Penedones,Paulos,Fitspatrick}. In order to examine more about the meaning of Eqs.~(\ref{eq13}) and (\ref{eq14}) with respect of this conjugate relation, we consider the inverse Mellin transformation of the correlator $\rho_{\eta}(l)=Al^{-\eta}\Gamma(\eta-\eta^{\prime})$. For this purpose, we analytically continue $\rho_{\eta}(l)$ into a holomorphic function by using the complex variable $\eta$.

\begin{figure}[htbp]
\begin{center}
\includegraphics[width=8.5cm]{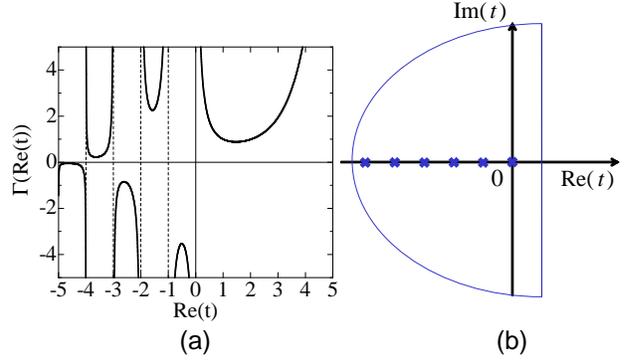}
\end{center}
\caption{(a) The gamma function along the real axis and (b) Poles of the complex gamma function and the semi-circle contour added to the Bromwich path.}
\label{snct67fig1}
\end{figure}

The inverse Mellin transformation is represented by the so-called Bromwich integral as
\begin{eqnarray}
{\cal R}(l,z)=\lim_{p\rightarrow\infty}\frac{1}{2\pi i}\int_{c-ip}^{c+ip}z^{-\eta}\rho_{\eta}(l)d\eta.
\end{eqnarray}
This integral is well-defined, when we find parameters $a$ and $b$ so that
\begin{eqnarray}
\int_{0}^{\infty}|{\cal R}(l,z)|z^{{\rm Re}(\eta)-1}dz<\infty, \label{eq18}
\end{eqnarray}
for $a<{\rm Re}(\eta)<b$. Then, there exists a parameter $c$ with the condition $a<c<b$. In the present case, it is enough to take $a=0$. By adding some integral contour to the Bromwich path, we pick up the pole of the gamma function in the density matrix $\rho_{\eta}(l)$ for this calculation. Since the gamma function for the complex number $t$ is represented by the infinite product formula
\begin{eqnarray}
\Gamma(t)&=&\lim_{n\rightarrow\infty}\frac{n!n^{t}}{t(t+1)(t+2)\cdots(t+n)} \\
&=&\frac{1}{t}\prod_{n=1}^{\infty}\frac{\left(1+\frac{1}{n}\right)^{t}}{1+\frac{t}{n}},
\end{eqnarray}
this has the poles on zero and negative intergers. The residues are given by
\begin{eqnarray}
{\rm Res}(\Gamma,-n)=\lim_{t\rightarrow -n}(t+n)\Gamma(t)=\frac{(-1)^{n}}{n!},
\end{eqnarray}
with $n=0,1,2,...$. This means
\begin{eqnarray}
{\cal R}(l,z) &=& \frac{1}{2\pi i}\oint_{C}z^{-\eta}\frac{A}{l^{\eta}}\Gamma\left(\eta-\eta^{\prime}\right)d\eta \nonumber \\
&=& \sum_{n\ge 0}\frac{A}{(lz)^{\eta}}{\rm Res}(\Gamma,-n)\delta\left(\eta-\eta^{\prime}+n\right) \nonumber \\
&=& \frac{A}{(lz)^{\eta^{\prime}}}e^{-lz},
\end{eqnarray}
where the integral path $C$ encloses all poles by adding an infinitely large semi-circle to the Bromwich path (see Fig.~\ref{snct67fig1}). We know that this actually agrees well with Eq.~(\ref{eq13}). By summing up all possible poles of the gamma function, we realized the correct off-critical behavior of the SVD component ${\cal R}(l,z=\xi^{-1})$ characterized by the exponential decay with the finite correlation length $\xi$. Therefore, the conjugate axis $\eta$ of the RG flow parameter $z$ acts as a source of such decay.

Now we start with Eq.~(\ref{eq14}) for the inverse transformation. However, the abovementioned calculation is impossible, if we start with Eq.~(\ref{eq16}). This is because there is no pole on the complex $\eta$ space. Thus we think that the gamma-function regulator is necessary to realize the well-defined inverse transformation.

\subsection{Roles of the Warp Factor on Off-Critical Behaviors of the Two-Point Spin Correlator}

\begin{figure}[htbp]
\begin{center}
\includegraphics[width=8cm]{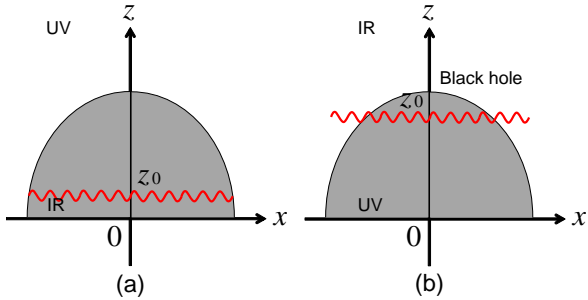}
\end{center}
\caption{Cut-off of the integral: (a) present classical/classical case and (b) BTZ black hole in the usual AdS/CFT (classical/quantum) correspondence.}
\label{snct67fig2}
\end{figure}

Up to now, we have examined the inverse Mellin transformation at the critical point by starting with $\rho_{\eta}(l)$. Furthermore, it is possible to extend this idea to the case away from the critical point. As already mentioned, we can introduce $\rho_{\eta}(l,z_{0})$ with the IR cut-off $z_{0}$ that is obtained from $\rho_{\eta}(l)$ by eliminating the large scale data. This situation is visualized in Fig.~\ref{snct67fig2}(a). The smaller $z$ value is related to the larger spatial scale contained in the original data. Away from the critical point, the larger scale data tends to disappear except for the background data of the snapshot.

According to the definition of the SVD spectrum in the original discrete representation, the spectrum always starts from $\lambda_{1}$ ($n=1$). Thus, we would like to keep the integration range of the parameter $z$ as $0\le z\le\infty$, even though the introduction of the IR cut-off $z_{0}$ is a very convenient method to represent the exponential damping factor away from the critical point. Furthermore, $\lambda_{1}$ becomes much larger than other $\lambda_{n}$ away from the critical point, and this feature is not simply represented by introducing the cut-off. Such situation can be phenomenologically represented by the warp factor like a black hole that terminates the RG flow space at $z=z_{0}$. We select the factor as $\sum_{k=0}^{\infty}(z_{0}/z)^{k}$, since this is equivalent to $\left(1-(z_{0}/z)\right)^{-1}$ for $z_{0}<z$ and this form is similar to the warp factor of the Schwarzschild black hole. We find
\begin{eqnarray}
\rho_{\eta}(l,z_{0}) &=& \sum_{k=0}^{\infty}\int_{0}^{\infty}dz\left(\frac{z_{0}}{z}\right)^{k}\frac{Ae^{-zl}}{(zl)^{\eta^{\prime}}}z^{\eta-1} \nonumber \\
&=& \frac{A}{l^{\eta^{\prime}}}\sum_{k=0}^{\infty}z_{0}^{k}\int_{0}^{\infty}dz e^{-zl}z^{\eta-\eta^{\prime}-k-1} \nonumber \\
&=& \frac{A}{l^{\eta}}\sum_{k=0}^{\infty}\left(z_{0}l\right)^{k}\int_{0}^{\infty}dx e^{-x}x^{\eta-\eta^{\prime}-k-1} \nonumber \\
&=& \frac{A}{l^{\eta}}\sum_{k=0}^{\infty}\left(z_{0}l\right)^{k}\Gamma(\eta-\eta^{\prime}-k). \label{eq29}
\end{eqnarray}
Here, we need to define $\Gamma(\eta-\eta^{\prime}-k)$ by analytic continuation. Let us further transform Eq.~(\ref{eq29}). According to the following property of the Gamma function
\begin{eqnarray}
\Gamma(t)=\frac{\Gamma(t+1)}{t},
\end{eqnarray}
the abovementioned result in the case of $\eta^{\prime}\simeq\eta$ can be transformed into
\begin{eqnarray}
\rho_{\eta}(l,z_{0}) &\simeq& \frac{A}{l^{\eta}}\sum_{k=0}^{\infty}\left(z_{0}l\right)^{k}\frac{\Gamma(\eta-\eta^{\prime})}{(-1)^{k}k!} \nonumber \\
&=& \frac{A}{l^{\eta}}e^{-z_{0}l}\Gamma(\eta-\eta^{\prime}).
\end{eqnarray}
We find that the result matches well with the two-point spin correlator away from the critical point (we must use Eq.~(\ref{eq29}) when considering the inverse transformation).

\section{Discussion}

Let us finally discuss some implications of the present results to the holography concept such as the AdS/CFT correspondence. In particular, we would like to mention similarity of our results with the correspondence except for the difference associated with the UV/IR relation. Since the present work is based on the classical/classical correspondence, the UV/IR relation would be reversed in comparison with the standard AdS/CFT that is a kind of the quantum/classical correspondence. What we would like to argue about the similarity is that the SVD index is a kind of the RG flow parameter.

Based on the above prerequisite, we should remark two important aspects. The first one is the appearance of the warp factor $(1-z_{0}/z)^{-1}$ in Eq.~(\ref{eq29}) when we consider the inverse Mellin transformation of the off-critical behavior. Going back to Eq.~(\ref{eq9}), we find
\begin{eqnarray}
\rho_{\eta}(l)=\int_{0}^{\infty}dz{\cal R}(l,z)z^{\eta-1}=\int_{0}^{\infty}\frac{dz}{z}{\cal R}(l,z)z^{\eta},
\end{eqnarray}
and this simple transformation indicates that the factor $dz/z$ represents the scale invariance of the hyperbolic geometry. We expect that the off-critical feature should be described by the black hole in quantum cases. For instance, the BTZ black hole geometry in $2+1$ dimension is given by the following metric
\begin{eqnarray}
ds^{2}=\frac{l^{2}}{z^{2}}\left(-f(z)dt^{2}+\frac{dz^{2}}{f(z)}+dx^{2}\right),
\end{eqnarray}
where the warp factor $f(z)$ is defined as
\begin{eqnarray}
f(z)=1-\left(\frac{z}{z_{0}}\right)^{2}.
\end{eqnarray}
The event horizon is located at $z=z_{0}$. Then, the truncation of the space by $f(z)$ occurs at $z\ge z_{0}$, as shown in Fig.~\ref{snct67fig2}(b). On the other hand, the warp factor in Eq.~(\ref{eq29}) terminates the RG flow for the region of $z<z_{0}$. Thus, the presence of the truncation or the termination for the flow parameter axis is quite similar in both classical and quantum cases, although the UV/IR region is reversed. This would be a strong indication of the similarity.

The second aspect is about the relation between the correlation length and the SVD index. This relationship has been precisely examined in terms of the matrix product state (MPS) formulation of correlated systems. Let us introduce the uniform MPS defined by
\begin{eqnarray}
\left|\psi\right>=\sum_{\left\{s_{j}\right\}}{\rm tr}\left(A[s_{1}]\cdots A[s_{N}]\right)\left|s_{1}\right>\otimes\cdots\otimes\left|s_{N}\right>,
\end{eqnarray}
where each matrix has $\chi\times\chi$ dimension. The best $\chi$ value is determined so that the entanglement entropy shows correct scaling behavior for a given model Hamiltonian. The two-point correlator for two local bosonic operators $O_{j}$ and $O_{j+l}$ (we can also define the fermionic case with some minor modification) is given by
\begin{eqnarray}
C(l)=\frac{\left<\psi\right|O_{j}O_{j+l}\left|\psi\right>}{\left<\psi|\psi\right>}=\frac{{\rm tr}\left(\hat{O}E^{l-1}\hat{O}E^{N-l-1}\right)}{{\rm tr}\left(E^{N}\right)},
\end{eqnarray}
where $\hat{O}$ and $E$ are respectively defined by
\begin{eqnarray}
\hat{O}=\sum_{s^{\prime},s}O_{s^{\prime},s}A^{\ast}[s^{\prime}]\otimes A[s],
\end{eqnarray}
and
\begin{eqnarray}
E=\sum_{s}A^{\ast}[s]\otimes A[s].
\end{eqnarray}
The MPS implies that in general a correlator takes the analytical form
\begin{eqnarray}
C(l)=\sum_{i=1}^{\chi^{2}}\alpha_{i}\lambda_{i}^{l}=\sum_{i=1}^{\chi^{2}}\alpha_{i}e^{-l/\xi_{i}},
\end{eqnarray}
where $\lambda_{i}$ is the eigenvalue of the $\chi^{2}\times\chi^{2}$ matrix $E$ and
\begin{eqnarray}\xi_{i}=-\frac{1}{\ln|\lambda_{i}|}.
\end{eqnarray}
By using the above formulation, it is important to examine the $\chi$ dependence on $\xi$, since $\chi$ is closely related to the RG flow parameter $z$. The examination has been done, and we summarize the results in the following for their comparison with the present result. 

In an earliar DMRG calculation~\cite{Andersson}, it was found that the particle-hole correlation in the 1D free fermion model (the central charge $c=1/2$) is given by
\begin{eqnarray}
\xi_{ph}\simeq -\frac{1}{\ln\left| 1-k\chi^{-\beta}\right|}\simeq \frac{1}{k}\chi^{\beta},
\end{eqnarray}
where $|\lambda_{ph}|\simeq 1-k\chi^{-\beta}$. The numerical fit of this scaling suggests $\beta\simeq 1.3$ and $k\simeq 0.45$. Furthermore, more precise analysis based on CFT tells us that the MPS with a finite dimention $\chi$ can approximately represent the non-local correlation scaled by
\begin{eqnarray}
\xi=\chi^{\kappa},
\end{eqnarray}
where $\kappa$ is the so-called the finite entanglement scaling exponent defined by
\begin{eqnarray}
\kappa=\frac{6}{c\left(\sqrt{12/c}+1\right)}, \label{eq43}
\end{eqnarray}
with the central charge $c$~\cite{Tagliacozzo,Pollmann}. In the Heisenberg model case $c=1$, we obtain $\kappa\simeq 1.344$ consistent with the abovementioned $\beta$ value. This consistency would be related to the fact that there are two different carriers in the analysis of the $\beta$ value. In both cases, $\xi$ increases as $\chi$ increases. These results also represent close connection between $\xi$ and $\chi$, and the relation is opposite to the present classical case $\xi^{-1}=z^{\kappa}$ where $z$ corresponds to $\chi$. The Ising model case ($c=1/2$), $\kappa$ in Eq.~(\ref{eq43}) is very close to $2$. This result may indicate that the functional form is $f(z)\propto e^{-l z^{2}}$ in the present classical case.

\section{Summary}

Summarizing, we defined the CSVD in the 2D classical Ising model, and examined its inverse Mellin transformation. We found that the continuous limit of the SVD index really corresponds to the flow parameter of the holographic RG in the sense that in this extended space the two-point correlator tends to be deformed so that the correlation length decreases as we go along this new axis. We also examined the effect of the IR cut-off on the exponential damping factor of the correlator away from the critical point. The present approach is based on the classical/classical correspondence. Thus, we found that the UV/IR relation is reversed in comparison with the AdS/CFT correspondence.

To proceed detailed analysis in this scale/Mellin space approach, we need to compare the present results with numerical data. This is because the present phenomenological theory still contains two adjustable parameters $\eta^{\prime}$ and $\kappa$. However, those numerical works are not easy, since we need to examine all of the matrix elements in the SVD. That would be the future important work. In the present work, we have examined the holographic RG data of the classical system. It is an interesting open question to examine the quantum case in close connection to the AdS/CFT correspondence.

HM acknowledges Ching Hua Lee for his helpful comments on the manuscript. This work was supported by JSPS Kakenhi Grant No.15K05222 and No.15H03652.

\end{document}